\def\arcdegg{\hbox{$^\circ$}}
\newcommand{\micron}{\,\hbox{$\mu$m} }
\newcommand{\micronn}{\,\hbox{$\mu$m}}
\def\etal{ {\em et~al.\/}\thinspace}
\def\arcsec{\hbox{$^{\prime\prime}$} }

\def\msun{\hbox{\,M$_\odot$}}

\def\lsun{\hbox{\,L$_\odot$}}

\def\rstar{\,R$_\star$}

\documentstyle[11pt,aaspp4,epsfig,flushrt]{article}
\lefthead{Monnier et al.}
\righthead{Mid-IR interferometry}

\begin{document}

%____________________________________TITLE PAGE___________________________
\title{Mid-infrared interferometry on spectral lines: 
\\
II. Continuum (dust) emission around IRC~+10216 and VY~CMa}
\author{J. D. Monnier\altaffilmark{1}, W. C. Danchi\altaffilmark{2}, 
D. S. Hale, E. A. Lipman\altaffilmark{3}, P. G. Tuthill\altaffilmark{4}, and 
C. H. Townes}

\altaffiltext{1}{Current Address: Smithsonian Astrophysical Observatory MS\#42,
60 Garden Street, Cambridge, MA, 02138}
\altaffiltext{2}{Current Address: NASA Goddard Space Flight Center,
Infrared Astrophysics, Code 685, Greenbelt, MD 20771}
\altaffiltext{3}{Current Address:
Laboratory of Chemical Physics, Building 5, Room 114, National Institutes 
of Health, Bethesda, MD, 20892-0520}
\altaffiltext{4}{Current Address: Chatterton Astronomy Dept, 
School of Physics, University of Sydney, NSW 2006, Australia}

\affil{Space Sciences Laboratory, University of California, Berkeley,
Berkeley,  CA  94720-7450}

%
%
%
%____________________________________ABSTRACT PAGE_________________________

\begin{abstract}

The U. C. Berkeley Infrared Spatial Interferometer has measured the
mid-infrared visibilities of the carbon star IRC~+10216 and the red
supergiant VY~CMa.  The dust shells around these sources have been 
previously shown to be time-variable, and these new data are used to
probe the evolution of the dust shells on a decade time-scale, complementing
contemporaneous studies at other wavelengths.
Self-consistent, spherically-symmetric models at maximum and minimum light
both show 
the inner radius of the IRC~+10216 dust shell to be much larger (150~mas) 
than that expected from
the dust condensation temperature, implying that dust production has
slowed or stopped in recent years.  
Apparently, dust does not form every pulsational cycle (638 days), and
these mid-infrared results are
consistent with recent
near-IR imaging which indicates little or no new dust production in
the last three years (Tuthill\etal 2000).  
Spherically symmetric models failed to fit 
recent VY~CMa data, implying that emission from
the inner dust shell is highly asymmetric and/or time-variable.

\end{abstract}

\keywords{instrumentation: interferometers,
techniques: interferometric,
stars: AGB and post-AGB, stars: circumstellar matter,
stars: winds}

%_______________________________________INTRODUCTION_______________________
\pagebreak
\section{Introduction}

Long-baseline infrared interferometry 
has been effective in
characterizing the {\em general} properties of dust shells around late-type
stars (e.g., \cite{danchi94}) for some years,  establishing 
that dust shells can be produced both
continuously (dust shell inner radii $\sim$2-3\,\rstar) and episodically
(dust shell inner radii $\gg$1\,\rstar) (e.g., \cite{dyck84};
\cite{danchi94}; \cite{bester96}).  However, recent
studies have also
shown that the dust shells around certain stars can change significantly
in the span of a few years
(e.g., \cite{bester96}; \cite{lopez97}; \cite{haniff98}),
challenging our understanding of the physics responsible for the
dust production and the mass-loss in general.

In order to understand what conditions 
precipitate the formation of stable, long-term dust and molecular
envelopes, regular observations of the dust shells around a number
of sources are required.  Monitoring the dust as it forms, is driven away by
radiation pressure, and ultimately replenished by fresh material,
promises to elucidate the intimate relationship between stellar
pulsations and dust production.   
High-resolution observations of the inner region of these dust shells are
easiest to interpret 
in the mid-infrared because of high dust opacities and scattered nebulosity
encountered in the visible and near-infrared.
In addition, interferometric observations allow for detection of
global asymmetries in the dust emission, but there have been few confirmed
reports of mid-infrared asymmetries around AGB stars (e.g., McCarthy 1979),
mostly because of limited spatial resolution.
Asymmetries in the mass-loss
flow are readily apparent in images of planetary nebulae (e.g., Sahai 1998), 
yet
it is still not clear how much asymmetry
begins during the AGB stage
or afterwards.

In this paper, we present visibility data 
from the Infrared Spatial Interferometer
(ISI) for carbon star IRC~+10216 and red supergiant 
VY~CMa.  It has been nearly a decade since
these stars have been observed at similar spatial resolution
(\cite{danchi94}), and these new data uniquely 
probe the temporal evolution of the dust shells over this time span.  In
addition, wide position angle coverage was obtained for observations of
VY~CMa, allowing the symmetry of its inner dust shell to be directly 
measured.  The next paper in this series (Monnier\etal 2000) 
utilizes the new dust shell models developed herein to interpret 
mid-infrared absorption lines from polyatomic molecules forming in the
dusty outflows.

\section{Methodologies}
\label{section:methodology}
A brief summary of the
observing and modeling methodologies for the Infrared Spatial
Interferometer will be presented in this section; 
further information can be found in
previous papers and publications, most notably Danchi\etal (1994) and
Hale\etal (2000).

\subsection{Observations}
The Infrared Spatial Interferometer (ISI) is a
two-element, heterodyne stellar interferometer operating at discrete
wavelengths in the 9--12~$\micron$ range and is located on Mt. Wilson,
CA.  The telescopes are each mounted within a movable semi-trailer and
together can currently operate at baselines ranging from 4 to 56~m.
Detailed descriptions of the apparatus and recent upgrades can be found
in Lipman (1998) and Hale\etal (2000).  System calibration
was maintained by observations of partially resolved
K giant stars $\alpha$~Tau and $\alpha$~Boo which have no 
known circumstellar dust.
This approach has proven to be successful at calibrating the
overall visibility data to a precision of about $\pm$5\%.  

By observing a target star throughout the night, 
visibility data are obtained at a variety of baselines, owing to the
changing projection of the telescope separation vector as the star
moves across the sky.  Data taken on various nights and with different
telescope spacings are collated and uncertainty estimates produced by
inspecting the internal scatter of the measurements at similar spatial
frequencies.  The sparse Fourier coverage afforded by the ISI's single
baseline and concomitant lack of Fourier phase (or closure phase)
measurements severely constrains the types of analyses that can be
pursued.  Thus aperture synthesis
techniques are of limited utility, 
and interpretation of the data relies largely on radiative transfer
calculations of simple dust shell models.

\subsection{Radiative Transfer Modeling}
\label{subsection:wolfire}
Physical parameters of dust shells around evolved stars can be
obtained through modeling, i.e. directly fitting the visibility data.
This allows additional observations, such as spectrophotometry or
interferometric work at other wavelengths, to be included as
constraints on the modeling.  The radiative transfer modeling code
used for this purpose is based on the work of Wolfire \& Cassinelli
(1986) and assumes the dust distribution to be spherically symmetric
(see Danchi\etal [1994] for a detailed description). Starting with the
optical properties and density distribution of the dust grains, this
code calculates the equilibrium temperature of the dust shell as a
function of the distance from the star, whose spectrum is assumed to
be a blackbody.  Subsequent radiative transfer calculations at 67
separate wavelengths allow the wavelength-dependent visibility curves,
the broadband spectral energy distribution, and the mid-infrared
spectrum all to be computed for comparison with observations.  The
8--12\,$\mu$m band is densely sampled, and has been useful for
comparing with UKIRT spectrophotometric monitoring (\cite{mgd98}).

The model simulates a star of radius
\rstar\, and temperature T$_{\rm eff}$ situated at the center of
the dust shell.  Because of high temperatures preventing dust
condensation, there is a dust-free spherical region extending from
the photosphere to
the dust shell inner radius, R$_{\rm dust}$.  For the simple models
under consideration here, the density distribution of dust is
constrained to follow a power law from R$_{\rm dust}$ to R$_{\rm
outer}$, at which point the dust density is assumed to drop to zero again.
Typically (and in all cases for this paper), {\em uniform outflow} is
assumed, implying the dust density falls off like $\rho\propto
R^{-2}$, and R$_{\rm outer}$ is taken to be large enough to contain
all significant thermal emission (in the mid-IR) of the dust.  The overall
density of the dust shell is parameterized by the optical depth $\tau$
at 11.15\,$\mu$m along a line-of-sight connecting the observer with the
star.

The code treats a distribution of dust sizes by calculating the dust
temperature as a function of grain size and dust type (e.g.  dirty
silicates, amorphous carbon, graphite), using the Mathis, Rumpl, \&
Nordsieck (1977) grain size distribution, where the grain size $a$
spans $0.01\,\micron$$<a<0.25\,\micron$ with number density $n
\propto a^{-3.5}$.
Dust opacities were calculated from the optical constants assuming
spheroidal Mie scattering using a method developed by Toon \& Ackerman
(1981).  
The warm silicate dust constants from Ossenkopf\etal (1992) and the
amorphous carbon (AC) dust constants from Rouleau \& Martin (1991) were
used for modeling oxygen-rich and carbon-rich dust shells
respectively.

\section{ISI Results and Modeling of Dust Shells}
\label{chapter:dust_models}
  New ISI visibility data for IRC~+10216 and VY~CMa presented here,
along with recent mid-infrared spectra, are
used to constrain radiative transfer models of the
circumstellar dust shells.  For any given source, it is difficult or
impossible to fit the vast array of multi-wavelength data available
using a simple model (see Monnier\etal [1997] and Groenewegen [1997]
for recent attempts).  This difficulty is largely due to the
overly-simple nature of the models (e.g., spherical symmetry), unknown
dust properties (e.g., grain size distribution, 
effects of large grains), and the lack of
coeval data for these pulsating stars.  Because we are partly
concerned with the dust shell as a mid-IR background
continuum source for molecular absorption, the modeling found herein
is focused on reproducing the observed mid-infrared properties in
Fall 1998 when molecular line data was collected (see Monnier\etal 2000).
Hence, the data and modeling results are presented for Fall
1998 separate from other epochs.
In order to implicitly include observations at other
wavelengths and from other epochs, 
some choices of model parameters have been guided by
the results of previous, more detailed modeling work, when
available.  

\subsection{IRC +10216}
\subsubsection{Introduction}
\label{section:10216_summary}
The long-period variable star IRC\,+10216 was discovered by Neugebauer
\& Leighton (1969) during the 2.2\,$\mu$m sky survey, and its optical
counterpart was shown to be extended and asymmetric (\cite{becklin69}).
It was soon recognized to be an evolved giant star surrounded by an
opaque envelope of dust, whose intense thermal emission makes it one
of the brightest mid-infrared sources outside the solar system.

Because of its high infrared brightness and the large number of molecules
found in its dense outflow, IRC~+10216 has become one of the most
studied objects in the galaxy; the Astrophysics Data System lists
nearly 1000 papers associated with this star.  In particular, the
large angular size of the dust shell, first established through lunar
occultation work (\cite{toombs72}), has made IRC~+10216 a favorite
target of interferometrists (see below).  

Spectral observations have revealed a carbon-rich envelope
(\cite{herbig70}) and subsequent workers have identified over 50 different
molecular species in the circumstellar environment
(\cite{glassgold98}), most in radio and millimeter transitions.
IRC\,+10216 has become prototypical of extreme evolved carbon stars, 
a rare stellar
specimen representing a brief, but important, phase of heavy mass-loss
in the late stages of stellar evolution.  The mass-loss rate has been
estimated through observations of J=2-1 line of CO and from
mid-infrared interferometric observations to be $\sim
3\times10^{-5}\msun yr^{-1}$ (\cite{knapp82};
\cite{danchi94}), assuming a distance of 135\,pc (\cite{groenewegen97}).
Recently, its inner dust shell has been imaged and shown to be highly
inhomogeneous and asymmetric (\cite{weigelt98a}; \cite{haniff98};
\cite{tuthill98b}; \cite{skinner98}; \cite{mythesis}; \cite{tuthill2000b}).
Observations of the outer envelope (e.g., \cite{bieging93}) indicate a
more or less spherically symmetric outflow, and so it has been
suggested that IRC~+10216 is presently undergoing a transformation
from a (spherically symmetric) red giant star to a (asymmetric)
planetary nebula.  If so, we are witnessing an extremely short-lived
phase and high spatial resolution observations are critical to
understanding the complicated processes at play.

While the inner region of the dust shell has already been shown to be
high clumpy (see references above) in the near-infrared and visible, 
the mid-infrared emission is expected to be more symmetrical 
because of enhanced emission from cooler dust
grains farther out in the flow, which contribute substantially to the total
mid-infrared flux. 
However, data from
the single baseline of the ISI cannot 
untangle complicated effects due to deviations from spherical
symmetry.  
A number of other authors have found that
spherically symmetric models of the dust shell around IRC~+10216 can
fit most of the spectral and imaging data, at least redward of
$\sim$2\,$\mu$m (e.g., \cite{ivezic96};
\cite{groenewegen97}).  At shorter wavelengths, scattering in
the inhomogeneous and clumpy inner dust shell dominate the appearance
(\cite{haniff98}; \cite{skinner98}) 
and spherically symmetric models begin to fail.  
While recognizing these uncertainties, in this section we
develop a spherically symmetric model of the dust
shell around IRC~+10216.  
 
\subsubsection{Data from Fall 1998}
The visibility curve of IRC~+10216 at 11.15\,$\mu$m
was measured
on 1998 Nov 18 (UT) with a 4-meter, East-West
baseline.  Earth rotation foreshorted the baseline and allowed a 
range of
spatial frequencies to be used. 
Details concerning data reduction and calibration
can be found in Hale\etal (2000) and references therein.

Table\,\ref{table:10216_99} and figure\,\ref{fig:vis_10216} show the
visibility data for IRC~+10216 calibrated using $\alpha$~Tau,
approximately a point source at the resolution of this experiment.
The ISI detects only radiation within the telescope primary beam,
FWHM$\sim$1.8$''$ for the 1.65\,m apertures.  This
significantly modifies the observed visibility curves from those
obtainable using an interferometer with smaller aperture telescopes.  
Active image
stabilization could not be used for obscured sources such as
IRC\,+10216 and allow the outer parts of the nebula to be detected as well.
This effect was included in the modeling described below by
multiplying the model emission profiles by an appropriately-sized
primary beam pattern centered on the central star.  The Gaussian 
FWHM used is reported in 
the parameter tables for all models.  Furthermore, the 
decorrelation
resulting from independent guiding errors in each telescope is
appropriately calibrated by observing the calibrator source under
similar observing conditions.

\begin{deluxetable}{cccc}
\tablecaption
{IRC +10216 Visibility Data from
November 1998 ($\lambda$=11.15\micronn) \label{table:10216_99}}
\tablehead{
\colhead{Spatial Frequency}  & \colhead{Position}     &
\colhead{Visibility} & \colhead{Visibility}\\
\colhead{(10$^5$ rad$^{-1}$)}&\colhead{ Angle (\arcdegg)} &         &
\colhead{Uncertainty}
}
\startdata
      2.19 & 110.5 &  0.314  &  0.017 \cr
      2.36 & 108.2 &  0.310  &  0.017 \\
      2.64 & 104.8 &  0.237  &  0.017 \\
      2.78 & 103.2 &  0.211  &  0.009 \\
      2.91 & 101.8 &  0.180  &  0.024 \\
      3.06 & 100.3 &  0.181  &  0.013 \\
      3.23 & 98.4  &  0.158  &  0.012 \\
      3.36 & 96.8  &  0.141  &  0.007 \\
      3.53 & 94.4  &  0.128  &  0.008 \\
      3.61 & 91.4  &  0.098  &  0.020 \\
\enddata
\end{deluxetable}

\subsubsection{The dust model}
\label{section:irc_dust}

\begin{deluxetable}{lcl}
\tablecaption
{IRC~+10216 Model Parameters for
Uniform Outflow Fit (November 1998 data) \label{table:par_10216_99}}
\tablehead{
\colhead{Parameter} & \colhead{Value} & \colhead{Comments}
}
\startdata
Distance (pc) & 135 & From Groenewegen (1997) \\
T$_{\rm eff}$ (K) & 2000 & From Groenewegen (1997) \\
Stellar Radius (mas) & 22.0 & \\
Luminosity (L$_\odot$) &  5892 & To fit mid-IR spectrum  \\
\hline
Dust Type & Amorphous Carbon & From Rouleau \& Martin (1991) \\
          & (type A-C) & \\
Dust Size & $n\propto a^{-3.5}$&  From Mathis, Rumpl, \& Nordsieck (1977) \\
\qquad Distribution && \\
R$_{\rm dust}$ (mas) & 150 & Inner radius of dust, \\
   & &  chosen to fit mid-IR visibility \\
$\tau$ at 11.15\micron & 0.66 & Optical Depth, \\
 & & chosen to fit mid-IR spectral slope \\
\hline
Effective Primary  & 3.00 &  \\
\qquad Beamwidth ($\arcsec$) & & \\
\enddata
\end{deluxetable}

\begin{figure}
\begin{center}
\centerline{\epsfxsize=4 in{\epsfbox{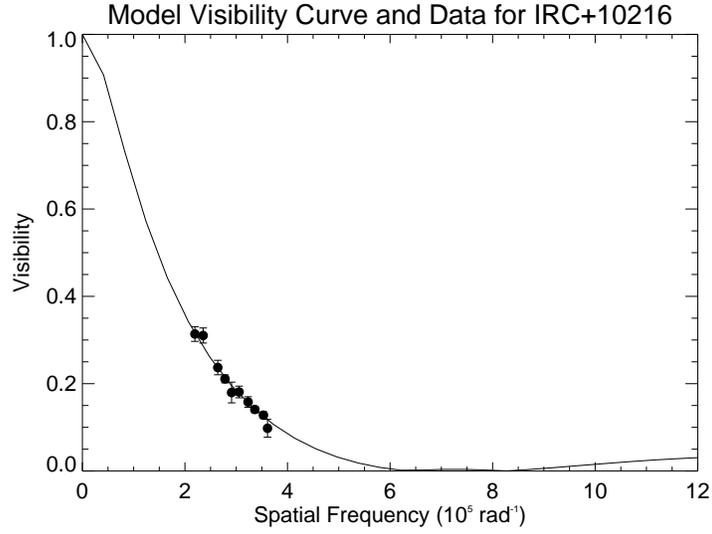}}}
\caption{The new mid-infrared visiblity data for IRC~+10216 in November 1998
and model results are shown.  The solid line is the model visibility curve base
d
on the parameters in Table\,\ref{table:par_10216_99}, assuming a
primary beam of 3\arcsec.
\label{fig:vis_10216}}
\end{center}
\end{figure}

\begin{figure}
\begin{center}
\centerline{\epsfxsize=3.5in{\epsfbox{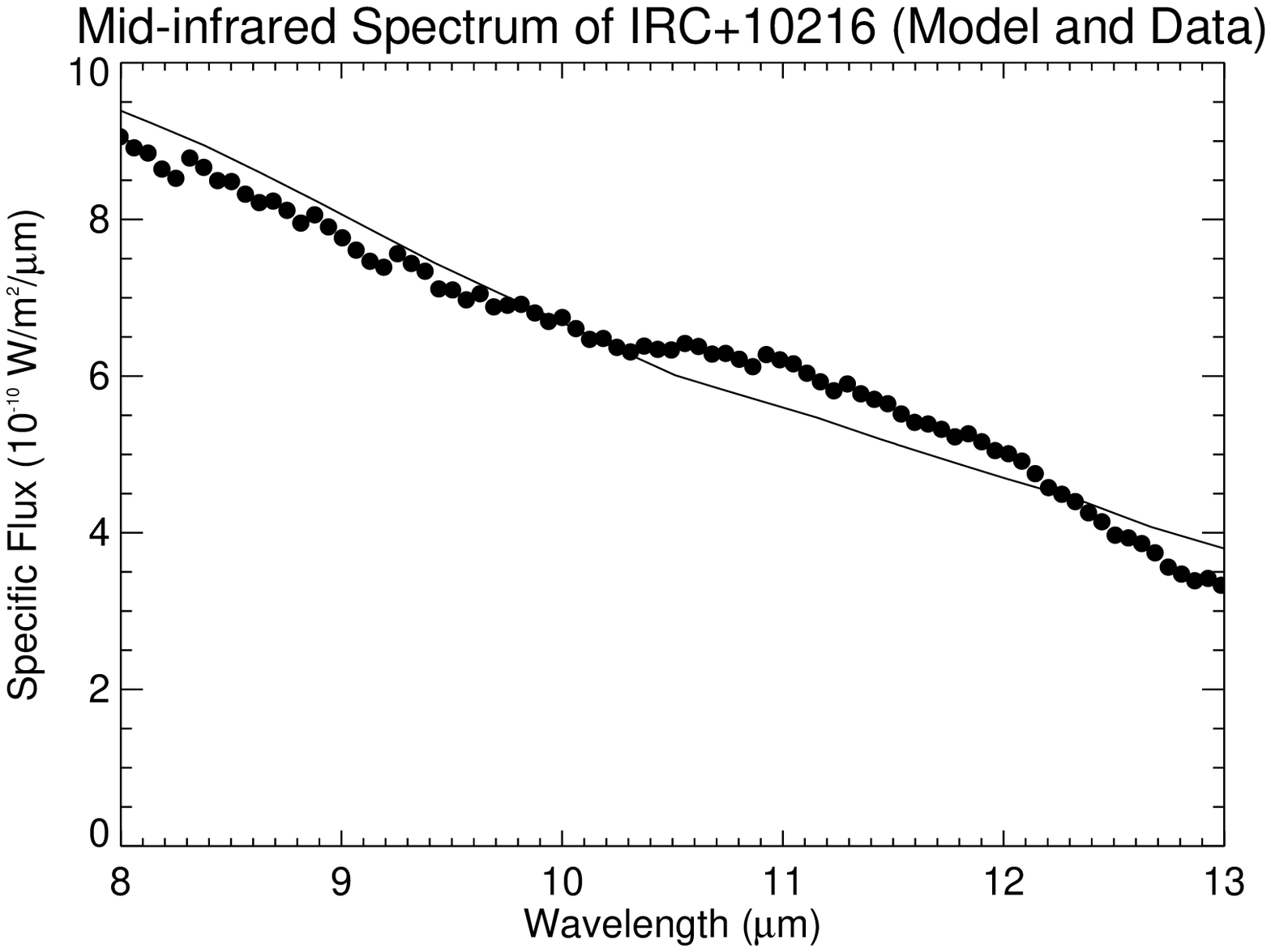}}}
\caption{The mid-infrared spectrum of IRC~+10216 taken on 1995 March 17 at a
phase similar to that of IRC~+10216 during 1998 November
interferometric observations (Monnier\etal 1998).
The solid line is the model spectrum based on the parameters in
Table\,\ref{table:par_10216_99}.
The dust mixture
used did not contain SiC, hence the spectral feature near 11.3\,$\mu$m
was not fitted here.
\label{fig:spec_10216}}
\end{center}
\end{figure}

The best-fitting model parameters for a uniform outflow dust shell
appear in table\,\ref{table:par_10216_99}, and were discussed 
in \S\ref{subsection:wolfire}.  The new ISI
visibility data from November 1998 and a mid-infrared spectrum taken at
pulsational phase 0.65 (Monnier\etal 1998) were
used to constrain the model fitting parameters.  This spectrum was
chosen to closely coincide with the pulsational phase of IRC~+10216
during the observation ($\phi_{\rm{IR}}\sim$0.76).  The data and model fits
can be found in figures\,\ref{fig:vis_10216} \& \ref{fig:spec_10216} and
are quite satisfactory, considering the simplicity of the dust density
distribution.  Note that the SiC feature near 11.3\,$\mu$m was not
included in the optical constants, resulting in a systematic
misfit in this spectral region.

The inner radius (150 mas) and mid-IR optical depth agree quite well
(within 10\%) with the dust shell parameters in
Groenewegen (1997), who recently modelled the source at maximum
light.  However, this inner radius is significantly 
larger than that deduced by models based on data taken approximately 
a decade ago, $\sim$75~mas 
deduced from earlier epochs of ISI data
(\cite{danchi94}) and $95\pm10$~mas from
near-infrared data (\cite{ridgway88}).
At 150~mas away from IRC~+10216 near minimum light, the
dust temperature is only about 860K, significantly below the
condensation temperature ($\sim$1400\,K).  
This can be explained if the dust was
initially created 5-10 years ago and has been flowing outward since
that time.  
This is also consistent with recent near-IR images of
the inner dust zone, which show a clumpy outwardly expanding dusty
environment with no obvious new dust production within the last 3
years (\cite{haniff98,weigelt98a,tuthill98b,mythesis,tuthill2000b}).  
This dust shell model will be used for subsequent modeling of
molecular absorption around IRC~+10216 (Monnier\etal 2000).

\subsubsection{Data from other epochs}

\begin{figure}
\begin{center}
\centerline{\epsfxsize=4 in{\epsfbox{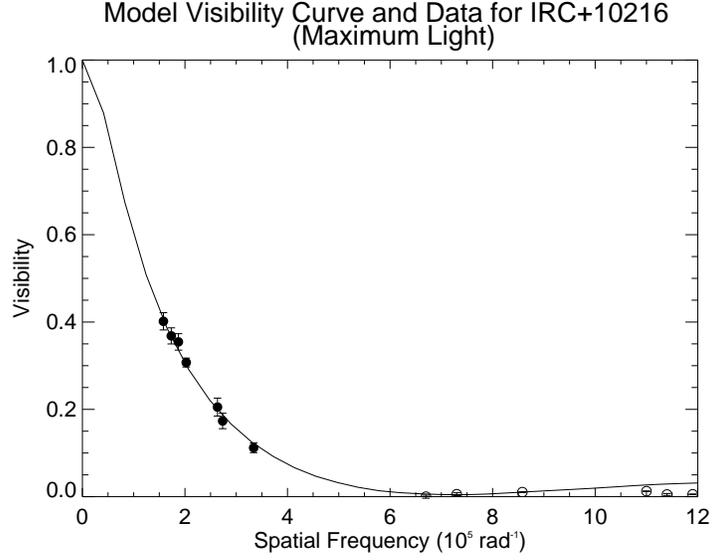}}}
\caption{Mid-infrared visibility data for IRC~+10216 taken on 1999 Apr 26-27
 (UT)
(filled circles, $\phi_{\rm{IR}}$ = 0.01) and
1997 Nov 1-23 (UT) (open circles,  $\phi_{\rm{IR}}$ = 0.16-0.19).
The model visibility curve is shown as a solid line assuming a
3\arcsec primary beam.
\label{fig:vis_10216_max}}
\end{center}
\end{figure}

\begin{deluxetable}{ccccc}
\tablecaption
{IRC +10216 Visibility Data from
November 1997 and April 1999 ($\lambda$=11.15\micronn)
\label{table:10216_other}}
\tablehead{
\colhead{Spatial Frequency}  & \colhead{Position}     & \colhead{Visibility}
 & \colhead{Visibility} & \colhead{Date of} \\
\colhead{(10$^5$ rad$^{-1}$)} & \colhead{Angle (\arcdegg)} &         & \colhead
{Uncertainty} & \colhead{Observation (UT)}
}
\startdata
1.58 & 64.7  & 0.402 & 0.020 & 1999 Apr 26-27 \\
1.73 & 67.6  & 0.368 & 0.018 & 1999 Apr 26-27 \\
1.87 & 69.4  & 0.355 & 0.019 & 1999 Apr 26-27 \\
2.03 & 72.3  & 0.307 & 0.010 & 1999 Apr 26-27 \\
2.64 & 79.6  & 0.205 & 0.021 & 1999 Apr 26 \\
2.74 & 80.2  & 0.173 & 0.018 & 1999 Apr 26 \\
3.34 & 87.1  & 0.112 & 0.011 & 1999 Apr 27 \\
6.7  & 326   & 0.0000& 0.0036& 1997 Nov 21 \\
7.3  & 317   & 0.0056& 0.0033& 1997 Nov 21 \\
8.6  & 299   & 0.0104& 0.0009& 1997 Nov 23 \\
11.0 & 347   & 0.0125& 0.0007& 1997 Nov 01 \\
11.4 & 338   & 0.0050& 0.0023& 1997 Nov 01 \\
11.9 & 332   & 0.0054& 0.0009& 1997 Nov 01 \\
\enddata
\end{deluxetable}

Special care was made to model the dust shell of IRC~+10216 during
Fall 1998 because spectral line data were also collected during this
epoch (Monnier\etal 2000).  However, additional data were collected on
16\,m and 9\,m baselines in early November 1997 and on a 4\,m baseline
in April 1999.  The (IR) pulsational phases for these two observations
were $\sim$0.18 and 0.01 respectively, and the data for these
observations can be found in table\,\ref{table:10216_other}.

Since both sets were taken near maximum light, the stellar luminosity
(15000~\lsun) and effective temperature (2000~K) from Groenewegen
(1997) were used for modeling the stellar output.  However, the dust
shell model developed to fit the Fall 1998 data was completely
unmodified (see table\,\ref{table:par_10216_99} for parameters).  The
additional luminosity decreased visibilities somewhat from Fall 1998
model values, and the comparison to the Spring 1999 data is quite
favorable (see figure\,\ref{fig:vis_10216_max}).  This means that the
differences in the mid-infrared visiblities can be completely
explained by the varying luminosity of the underlying star, without
invoking new dust production or significant dust shell evolution
between epochs.  The disagreement at the longest baseline may be due
to inhomogeneities and asymmetries in the dust shell, reflecting very
high resolution structure not incorporated in spherically symmetric
models.

\subsection{VY CMa}

\subsubsection{Introduction}
\label{section:vycma_summary}

VY~CMa (spectral type M5eIbp) is a very unusual star, displaying
intense radio maser emission from a variety of molecules, strong dust
emission in the mid-infrared, high polarization in the near-infrared,
and large amplitude variability in the visible.  Many properties of VY~CMa
were recently reviewed in Monnier\etal (1999a), which included 
new near-IR imagery of its complicated circumstellar environment.

Using a distance estimate of 1.5~kpc based on work by
Lada \& Reid (1978) and observations of maser proper motions
(\cite{marvel96}; \cite{richards98}),  
Monnier\etal (1999a) estimated the true, obscuration-corrected
luminosity to be L$_\star \approx$2$\times 10^5\,$L$_\odot$.
The high
luminosity, coupled with an extremely low effective temperature
T$_\star \approx 2800\,$K (\cite{sidaner96}), suggests VY CMa is a
massive star (M$_\star \approx 25$M$_\odot$) on the verge of exploding
as a supernova (within $\sim$10$^4$ years, \cite{bt82}).  Another sign of
impending cataclysm is the extensive mass being lost by VY~CMa into an
optically thick circumstellar envelope.  The mass loss rate for this
star has been estimated using a variety of techniques (summarized in
Danchi\etal[1994]), yielding a most probable value of {$\dot{M}\approx$
2$\times10^{-4}\,$M$_\odot \,{\rm{yr}}^{-1}$}.

Previous observations of the circumstellar envelope at a variety of
wavelengths have shown evidence of significant asymmetries.  
Maser emission of SiO, H$_2$O, and OH show spatial and redshift
distributions not consistent with simple outflow geometries 
(e.g. \cite{bm79}; \cite{marvel96}; 
\cite{richards98}).   In addition, optical observers have noted 
VY~CMa's peculiar one-sided nebulosity for most of this century
(\cite{worley72}; \cite{herbig72}).
Indeed, recent visible images from the
Hubble Space Telescope (\cite{kw98,smith99}) and
near-IR adaptive optics observations (\cite{monnier99a}) show a
one-sided reflection nebula, with a complicated arrangement of
scattering features.  High values of near-infrared linear polarization
have been observed, resulting from small-scale asymmetries close
to the star itself (\cite{monnier99a}).
McCarthy (1979) reported asymmetry
in the mid-infrared emission of the dust shell, interpreting this as
evidence for thermal emission from a disk-like structure.  Another
important observational fact is that VY~CMa is an irregular variable
star showing 1-3~mag optical variation on the time scale of
$\sim$2000~days (\cite{marvel96}).  Other observations testifying to the
dynamic nature of this source are
recent changes in the
direction of the near-infrared polarization (\cite{maihara76})
and shape of the
mid-infrared spectrum (Monnier, Geballe, \& Danchi 1998, 1999)

\subsubsection{Data from Fall 1998}
The 11.15\,$\mu$m visibility curve of VY CMa was measured over a
two-day period, from 1998 November 18 to 1998 November 19 (UT) with
the ISI in its 4-meter, East-West baseline configuration.  Details
concerning data reduction and calibration are identical to those for
IRC\,+10216.  

\begin{deluxetable}{cccc}
\tablecaption{
VY CMa Visibility Data from November
1998 ($\lambda$=11.15\micronn) \label{table:vycma_99}}
\tablehead{
\colhead{Spatial Frequency}  & \colhead{Position}    & \colhead{Visibility}
 & \colhead{Visibility}\\
\colhead{(10$^5$ rad$^{-1}$)}& \colhead{Angle (\arcdegg)} &         &
\colhead{Uncertainty} }
\startdata
      2.51 & 62.5 &   0.49  &  0.10   \\
      2.63 & 65.3 &   0.46  &  0.06   \\
      2.79 & 68.2 &   0.42  &  0.04   \\
      2.93 & 71.7 &   0.41  &  0.05   \\
      3.08 & 74.9 &   0.33  &  0.04   \\
      3.23 & 77.6 &   0.36  &  0.03   \\
\enddata
\end{deluxetable}

\subsubsection{The dust model}
\label{section:vyc_dust}

\begin{deluxetable}{lcl}
\tablecaption{
VY CMa Model Parameters for
Uniform Outflow Fit (November 1998 data)\label{table:par_vycma_99}}
\tablehead{
\colhead{Parameter} & \colhead{Value} & \colhead{Comments} }
\startdata
Distance (pc) & 1500 & From Lada \& Reid (1978) \\
T$_{\rm eff}$ (K) & 2700 & From Monnier\etal (1999a) \\
Stellar Radius (mas) & 10.0 & \\
Luminosity (L$_\odot$) &  5$\times$10$^5$ & Chosen to fit mid-IR spectrum \\
\hline
Dust Type & Silicates & From Ossenkopf, Henning, \& Mathis (1992) \\
Dust Size & $n\propto a^{-3.5}$  &  From Mathis, Rumpl, \& Nordsieck (1977) \\
\qquad Distribution && \\
R$_{\rm dust}$ (mas) & 50 & Dust inner radius, \\
&& Chosen to fit mid-IR visibility \\
$\tau$ at 11.15\micron & 2.41 & Optical depth, \\
&& Chosen to fit mid-IR spectral shape \\
\hline
Effective Primary  & 3.0 &  \\
\qquad Beamwidth ($\arcsec$) & & \\
\enddata
\end{deluxetable}

\begin{figure}
\begin{center}
\centerline{\epsfxsize=4 in{\epsfbox{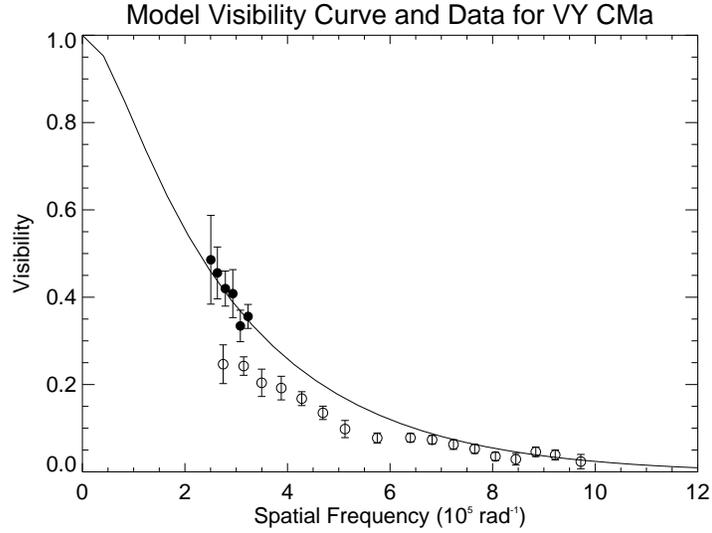}}}
\caption{The mid-infrared visibility data for VY CMa and the
model results are shown.  The model visibility curve (solid line)
is based on data obtained in
Fall 1998 (solid circles), while Fall 1997 data are represented by
open circles (see \S\ref{section:vy_other} for discussion of
discrepancy).  See Table\,\ref{table:par_vycma_99} for model
parameters.
\label{fig:vis_vycma}}
\end{center}
\end{figure}

\begin{figure}
\begin{center}
\centerline{\epsfxsize=3.5in{\epsfbox{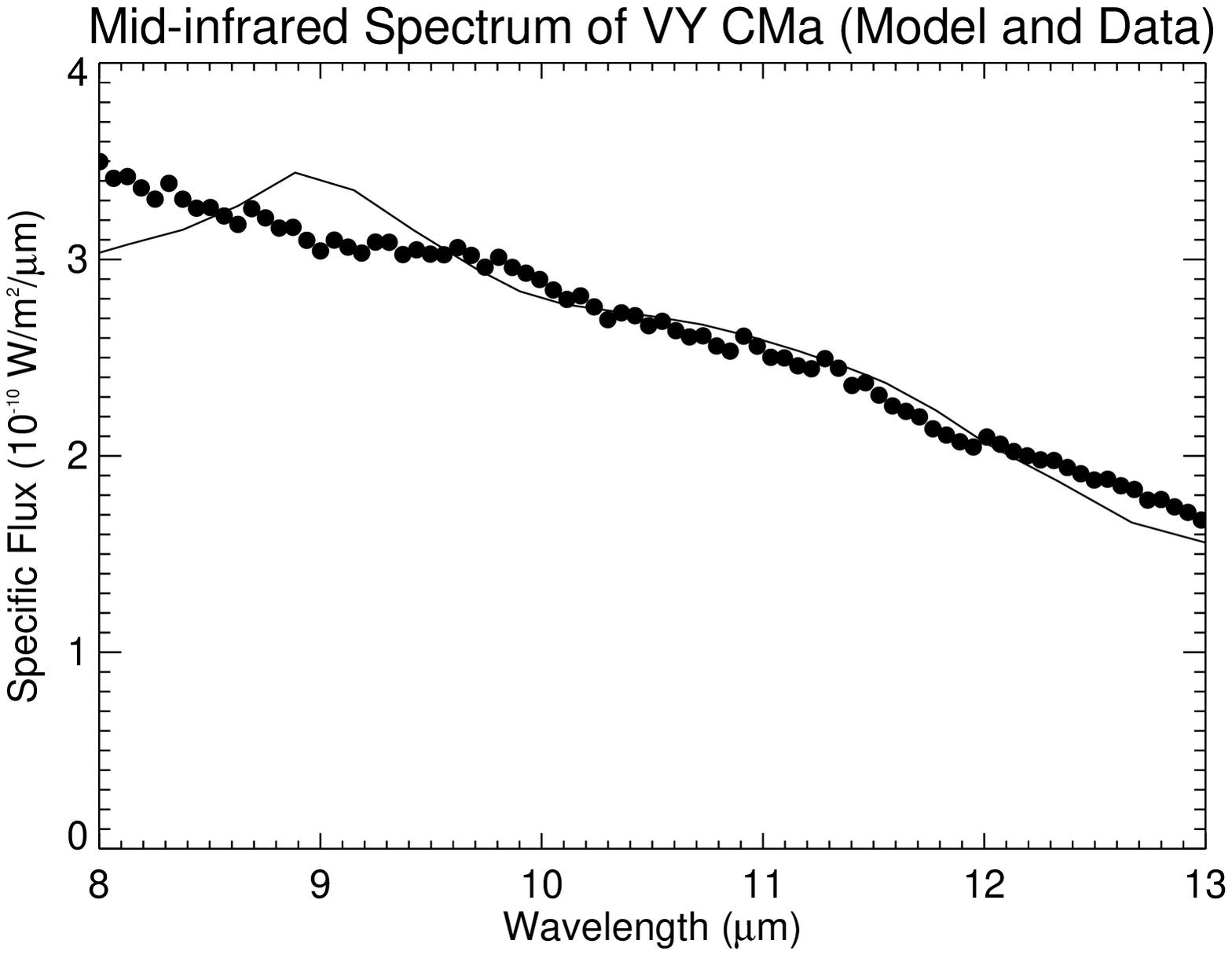}}}
\caption{A typical mid-infrared spectrum of VY CMa (from 1995 March 17;
Monnier\etal [1998]) and the
model fit (solid line).  See Table\,\ref{table:par_vycma_99}
for more details on model parameters.
\label{fig:spec_vycma}}
\end{center}
\end{figure}

Despite the evidence for asymmetries in the inner dust shell,
spherical symmetry was adopted for modeling the average properties of
the mid-IR emission.  The best-fitting model parameters for a uniform
outflow dust shell appear in table\,\ref{table:par_vycma_99}.  The new
ISI visibility data from November 1998 and a mid-infrared spectrum
(Monnier\etal 1998) were used to constrain the model fitting
parameters.  VY CMa does not have a well-defined pulsational
period and because we have no good estimate of its flux level
in November 1998, a
mid-IR spectrum of intermediate brightness (1995 Mar 17) was used.
The data and model fits can be found in figures\,\ref{fig:vis_vycma}
\& \ref{fig:spec_vycma} and are in good agreement.  Note that the
model silicate feature near 9.7\,$\mu$m 
appears with a decreased peak due to self-absorption,
although not as flat as the observed spectrum.

   The inner radius (50\,mas) and mid-IR optical depth agree well with
previous modeling based on mid-infrared interferometric measurements
(\cite{danchi94}).  According to the current model, the dust
temperature at this inner radius is about 1300 K, which is roughly the
expected condensation temperature for silicate dust grains.  Note that
this inner radius is in striking disagreement with the modeling
results of Le Sidaner (1996) whose fitting procedure only considered
the broadband spectral energy distribution (SED).  While their best
fitting models indicate a dust shell inner radius of 120 mas, the
authors noted that the VY CMa results were ``certainly the least
satisfactory in [their] work.''  This illustrates the importance of
high angular resolution observations for properly constraining
radiative transfer models.

Because the dust shell around this star appears asymmetric on all
observed scales (10-10000 AU; e.g., \cite{monnier99a}), this simple
spherically symmetric model should be considered a
rough approximation of the true
dust shell; we attempt here only to represent some kind of ``average'' dust
density distribution.  This has proven adequate for interpreting the
spectral line results discussed in the next paper of this series
(Monnier\etal 2000).

\subsubsection{Data from other epochs}
\label{section:vy_other}
VY CMa was also observed by the ISI in October 1997 with a 16m
baseline.  This configuration allowed the visibility to be sampled
at projected baselines from about 
3 to 10 meters.  These data have been plotted along with Fall
1998 data on figure\,\ref{fig:vis_vycma} (numerical values can
be found in table\,\ref{table:vycma_other}).  
Significant differences in the short baseline data at the two epochs
can be seen (at $\sim 3 \times 10^{5}$\,rad$^{-1}$),
plausibly arising from changes in the dust shell between observations,
deviations from circular symmetry, or anomalous miscalibration at low
elevation.  These possibilities will be discussed in turn.

Near-constancy in the near-infrared brightness distribution from
January 1997 to January 1999 (Monnier\etal 1999a; recent unpublished
data) suggests that no significant changes in the dust shell 
occurred between the ISI observations of 
October 1997 and November 1998.  Interestingly, the
short baseline observations in 1997 and 1998 were sampling nearly
orthogonal position angles on the sky (see
tables\,\ref{table:vycma_99} \& \ref{table:vycma_other}, where angles
are measured in degrees East of North). Hence, the visibility
differences could be due to dust shell asymmetry
(e.g., NW-SE elongation).  Unfortunately, VY~CMa was at 
unavoidably low sky elevations during short baseline observations on the
16m baseline.  While accurate calibrations were generally attainable in
such cases, it is possible that anomalous atmospheric conditions
at low elevation could
have caused the lower visibility observed, mimicking the aforementioned
dust shell asymmetry.

An asymmetric dust distribution is suggested by high
resolution, near-infrared and visible imaging of VY~CMa's circumstellar 
environment.
A NW-SE elongation direction roughly coincides with the ``equatorial
plane'' of enhanced mass-loss inferred to exist from the morphology of
the reflection nebula (\cite{kw98,monnier99a}), but is not in
agreement with an earlier mid-IR measurement reported by McCarthy\etal
(1980).  Diffraction-limited imaging with 8\,m-class telescopes in 
the mid-infrared should resolve these ambiguities in
interpretation.

\begin{deluxetable}{cccc}
\tablecaption{
VY CMa Visibility Data
from 1997 October 20 ($\lambda$=11.15\micronn) \label{table:vycma_other}}
\tablehead{
\colhead{Spatial Frequency } & \colhead{Position}     & \colhead{Visibility}
 & \colhead{Visibility}\\
\colhead{(10$^5$ rad$^{-1}$)}& \colhead{Angle (\arcdegg)} &         &
\colhead{Uncertainty} }
\startdata
  2.7 & 330 & 0.247 & 0.044 \\
  3.1 & 320 & 0.242 & 0.021 \\
  3.5 & 314 & 0.204 & 0.031 \\
  3.9 & 310 & 0.192 & 0.027 \\
  4.3 & 306 & 0.168 & 0.016 \\
  4.7 & 304 & 0.135 & 0.015 \\
  5.1 & 301 & 0.098 & 0.020 \\
  5.8 & 299 & 0.077 & 0.012 \\
  6.4 & 297 & 0.078 & 0.010 \\
  6.8 & 296 & 0.073 & 0.010 \\
  7.2 & 296 & 0.063 & 0.011 \\
  7.7 & 295 & 0.053 & 0.010 \\
  8.1 & 295 & 0.035 & 0.009 \\
  8.5 & 295 & 0.029 & 0.013 \\
  8.8 & 294 & 0.046 & 0.011 \\
  9.2 & 294 & 0.039 & 0.011 \\
  9.7 & 294 & 0.024 & 0.016 \\
\enddata
\end{deluxetable}

\section{Conclusions}
We have presented new visibility data and models of the dust shells
for the carbon star IRC~+10216 and
the red supergiant VY~CMa.  Spherically symmetric, uniform outflow models
were adequate to fit most of the mid-infrared properities of these
sources.  We find that the inner radius of the IRC~+10216 dust shell is 
much larger than that expected if dust was continuously condensing out of
the gas; this is in contrast to earlier epochs which showed material 
much closer to the star.  From this, we 
conclude that little new dust has been produced
around IRC~+10216 during the last 5-10 years.
However, the dust shell around VY CMa appears very similar to that first
observed by the ISI around 1990, implying continuous production of dust 
over the intervening years.  These new data have enhanced 
position angle coverage, yielding evidence for a marked asymmetry of the 
inner dust shell of VY CMa, although
the data are also consistent with pure time evolution.  The dust envelopes
around these stars are sufficiently large that diffraction-limited 
observations using a mid-infrared instrument on an 8\,m-class telescope
should
resolve possible asymmetries and vastly improve our understanding of
these dust shells, as well as the physical causes for the development
of mass-loss asymmetry 
during the final stages of stellar evolution.

\acknowledgements
{
This work is a part of a
long-standing interferometry program at U.C. Berkeley, supported by
the National Science Foundation (Grant AST-9221105, AST-9321289,
and AST-9731625)
, the Office of Naval Research (OCNR N00014-89-J-1583), and NASA.}

\end{document}